\documentclass[pdflatex,sn-basic]{sn-jnl}

\usepackage{graphicx}
\usepackage{multirow}
\usepackage{amsmath,amssymb,amsfonts}
\usepackage{amsthm}
\usepackage{booktabs}
\usepackage[title]{appendix}
\usepackage{xcolor}
\usepackage{textcomp}
\usepackage{eurosym}      

\begin{document}

\title[Nowcasting Italian Municipal Income with Nightlights]{Nowcasting Italian Municipal Income with Nightlights: A Deep Learning Approach}

\author*[1]{\fnm{Massimo} \sur{Giannini}}\email{massimo.giannini@uniroma2.it}

\affil*[1]{\orgdiv{Department of Enterprise Engineering}, \orgname{University of Rome Tor Vergata}, \orgaddress{\street{Via del Politecnico, 1}, \city{Rome}, \postcode{00133}, \country{Italy}}}

\abstract{This paper assesses whether NASA Black Marble nightlight intensity can serve as an early indicator of annual taxable income at the Italian municipal level, where official data are released with a 12--18 month lag. Using a panel of 7{,}631 municipalities over 2012--2021, we compare four recurrent neural network architectures (LSTM, BiLSTM, GRU, Transformer) against six benchmarks: simple persistence, panel fixed effects, autoregressive distributed lag, and two spatial econometric specifications (SAR, Spatial Durbin) on a queen-contiguity matrix. Models are trained on 2012--2019 and evaluated out-of-sample on 2020--2021 with a cross-sectional Diebold--Mariano test. A single-layer GRU achieves a median forecast error of 1.07 million euros across the cross-section of municipalities --- approximately $4\%$ of the median municipal IRPEF income of 29 million euros --- statistically dominating every benchmark (DM $>4$ against persistence, $>40$ against spatial linear models, all $p<0.001$). Spatial models recover statistically significant spatial autocorrelation ($\rho \approx 0.71$) and a meaningful nightlight spillover ($\theta \approx 0.05$), but their forecasting gap with the GRU is virtually identical to that of spatially-naive linear specifications. We conclude that nightlights contain genuine predictive content for municipal income, but extracting it requires a model class flexible enough to capture cross-sectional heterogeneity and non-linearities that linear specifications, spatial or otherwise, cannot recover.}



\maketitle

\section{Introduction}
\label{sec:intro}

Timely measurement of subnational economic activity is a long-standing
challenge for fiscal policy, regional planning, and economic geography. In
Italy, the most granular and reliable measure of local economic activity
is the personal income tax base (\textit{reddito imponibile} from the
IRPEF declarations), available at the municipality level from the Ministry
of Economy and Finance. However, the data are released with a delay of
12 to 18 months: 2020 figures, for instance, were published in late 2021
and early 2022. This delay constrains the timeliness of regional economic
monitoring and limits the responsiveness of fiscal interventions, especially
during shock periods such as the COVID-19 pandemic when local economic
conditions diverged rapidly across the territory.

Satellite-based nightlight imagery has emerged in the last fifteen years as
a complementary high-frequency proxy for economic activity. NASA's Black
Marble product \citep{roman2018nasa} provides global monthly composites of
stable nighttime radiance at 500-meter resolution, with a release lag of
approximately three weeks. A growing body of work has documented robust
cross-sectional correlations between nightlight intensity and economic
aggregates at the country level \citep{henderson2012measuring,chen2011using},
the regional level \citep{donaldson2016view,bickenbach2016night}, and most
recently the municipal level \citep{fiaschi2024income}. This literature has
primarily focused on \emph{cross-sectional} relationships -- using
nightlights to compare income across spatial units in a given year --
rather than on the \emph{temporal} and \emph{predictive} properties of the
relationship.

This paper studies whether nightlights can serve as a leading indicator
for the temporal evolution of municipal income, providing nowcasts before
the official data release. We frame the problem as a panel forecasting
exercise on 7{,}631 Italian municipalities over 2012--2021, training models
on 2012--2019 and evaluating out-of-sample on 2020--2021 -- a test period
that includes the COVID-19 shock and its asymmetric local impact. Our central
question is whether deep learning architectures can extract predictive
content from nightlight series that conventional linear and spatial-linear
specifications miss.

\medskip

We make three contributions. First, we document that the nightlight--income
relationship at the Italian municipal level is strongly non-linear and
heterogeneous across the cross-section. Despite a high pooled correlation
of $0.92$ between annual nightlight intensity and IRPEF income -- driven by
cross-sectional variation in size and economic activity -- the
within-municipality temporal correlation has a median of only $0.10$, and
only $41\%$ of municipalities exhibit a temporal correlation above $0.30$.
This heterogeneity implies that any single linear specification, even one
with municipality-specific intercepts, will misspecify a substantial
fraction of the panel.

Second, we assess whether explicit modeling of spatial dependence improves
nowcasting performance. We estimate Spatial Autoregressive (SAR) and
Spatial Durbin (SDM) panel models with municipality fixed effects on a
queen-contiguity weights matrix derived from ISTAT shapefiles. We find
statistically significant spatial autocorrelation ($\rho = 0.71$) and a
positive spatial spillover from neighbouring nightlights to local income
($\theta = 0.05$ in the Durbin specification). However, modeling spatial
dependence does not narrow the forecasting gap relative to the deep
learning models: SAR and SDM achieve Diebold--Mariano statistics against
the GRU benchmark of approximately $+40$ ($p<0.001$), virtually identical
to those of spatially-naive linear specifications. To our knowledge, this
is the first evaluation of spatial econometric models against recurrent
neural network forecasters on a panel of this dimension.

Third, we conduct a horse race among recurrent architectures that addresses
methodological concerns identified in prior applications of deep learning
to economic forecasting. We use a panel-trained sequence-to-sequence setup;
for bidirectional architectures (BiLSTM, Transformer with self-attention)
we implement iterative one-step-ahead forecasting to avoid in-sample
smoothing; we standardize inputs using training-period statistics only to
prevent information leakage; and we evaluate forecast accuracy with the
\citet{pesaran2007simple} cross-sectional Diebold--Mariano test, appropriate
for short-$T$ panels. The simplest architecture -- a single-layer GRU --
emerges as the best performer, consistent with parsimony principles in
panel forecasting documented by \citet{coulombe2022machine} and
\citet{medeiros2021forecasting}.

\medskip

Our main result is that the GRU produces a median forecast error of
approximately $1.07$ million euros across the cross-section of
municipalities --- corresponding to roughly $4\%$ of the median
municipal IRPEF income of $29$ million euros, an $18\%$ improvement
over a naive persistence benchmark and a $38\%$ improvement over the
best linear specification. The GRU statistically dominates all nine alternatives,
including the spatially-augmented benchmarks, and its dominance is
uniformly distributed across the cross-section: only $18\%$ of total mean
squared error originates from the top $1\%$ of municipalities. We conclude
that nightlights contain genuine predictive content for municipal income at
annual frequency, but extracting this content requires a model class
flexible enough to capture cross-sectional heterogeneity and non-linearity
in the conditional mapping.

The remainder of the paper is organized as follows.
Section~\ref{sec:data} describes the data sources and presents diagnostic
statistics on the integration order and correlation structure of the panel.
Section~\ref{sec:method} details the forecasting setup, the four recurrent
architectures, and the six benchmark models, including the spatial
specifications. Section~\ref{sec:results} presents the main empirical
results. Section~\ref{sec:robustness} discusses robustness checks and
limitations. Section~\ref{sec:conclusions} concludes.


\section{Data}
\label{sec:data}

We construct an annual panel of 7{,}645 Italian municipalities (\textit{comuni})
covering the period 2012--2021. The two key variables are the spatial
intensity of nighttime radiance, derived from NASA Black Marble satellite
products, and the personal income tax base (\textit{reddito imponibile
IRPEF}) released annually by the Italian Ministry of Economy and Finance.
Section~\ref{sec:robustness} and the spatial econometric robustness analysis
restrict the panel to the 7{,}631 municipalities for which queen-contiguity
spatial neighbours can be identified after iterative removal of isolated
small islands; the deep learning models and statistical benchmarks use the
full sample.

\subsection{Nightlight intensity}
\label{sec:data_nl}

Nightlight data come from NASA's Black Marble suite \citep{roman2018nasa},
specifically the VIIRS Day/Night Band monthly composite product (VNP46A3),
which corrects raw radiance for atmospheric, lunar, and seasonal
contamination and is delivered on a $15$~arc-second grid (approximately
$500$~metres at the equator). For each Italian municipality we extract the
mean radiance over the polygons defined by the 2019 ISTAT administrative
boundaries (\texttt{Com01012019\_g\_WGS84}), then aggregate the monthly
series to annual frequency by summation, obtaining a panel matrix
$\mathrm{NL}_{i,t}$ for $i=1,\dots,N$ municipalities and $t=2012,\dots,2021$.
The choice of annual aggregation matches the temporal granularity of the
income data and is consistent with most of the empirical literature on
nightlights and economic activity (\citealt{henderson2012measuring};
\citealt{donaldson2016view}). Robustness checks reported in
Section~\ref{sec:robustness} verify that intra-annual disaggregation of the
nightlight series (quarterly inputs) does not alter the main results.

\subsection{Municipal income}
\label{sec:data_income}

The target variable is the aggregate IRPEF taxable income at the municipal
level, sourced from the open data portal of the Department of Finance
(MEF). For each fiscal year $t$, the value $y_{i,t}$ represents the sum
of declared taxable income across all resident taxpayers in municipality
$i$, expressed in current euros. Data are released with a delay of
approximately 12 to 18 months: 2020 figures, for instance, were published
in late 2021 to early 2022. We work with raw nominal values without
deflating, since the test period is short (two years) and a common GDP
deflator across all municipalities would not affect the relative ranking
of forecasts.

For seven municipalities affected by administrative mergers between 2012
and 2021, we map the historical income figures to the 2019 boundary
definition by summing the income of pre-merger predecessor municipalities,
ensuring temporal consistency of the panel.\footnote{The mapping uses the
ISTAT correspondence tables for municipal codes available at
\url{https://www.istat.it/it/archivio/6789}.}

\subsection{Descriptive statistics and panel structure}
\label{sec:data_diagnostics}

Table~\ref{tab:descriptives} reports descriptive statistics for nightlight
intensity and IRPEF income, pooled across municipalities and years.

\begin{table}[h]
\centering
\small
\caption{Descriptive statistics, 7{,}645 municipalities $\times$ 10 years (2012--2021)}
\label{tab:descriptives}
\begin{tabular}{lrrrrrrr}
\toprule
 & Mean & SD & P5 & Median & P95 & Min & Max \\
\midrule
NL intensity (raw) & 9{,}529.44 & 34{,}822.18 & 599.06 & 3{,}758.16 & 32{,}020.27 & 60.65 & 2{,}581{,}092.73 \\
log(NL+1) & 8.29 & 1.22 & 6.40 & 8.23 & 10.37 & 4.12 & 14.76 \\
IRPEF (M\euro) & 103.83 & 743.93 & 3.26 & 29.16 & 312.00 & 0.38 & 50{,}214.04 \\
\bottomrule
\end{tabular}

\medskip
\footnotesize
\textit{Notes.} NL intensity is the annual sum of monthly Black Marble VNP46A3
mean radiance values within each municipal polygon. IRPEF is the aggregate
declared taxable income from personal income tax declarations. Both variables
are expressed at the municipality--year level; statistics are pooled across
$7{,}645 \times 10 = 76{,}450$ observations.
\end{table}

The panel exhibits substantial cross-sectional heterogeneity in scale: the
ratio between the largest (Roma Capitale, with annual IRPEF of approximately
50 billion euros) and the smallest municipality by income exceeds five
orders of magnitude.\footnote{We do not exclude small municipalities or apply
size-based weighting in the main analysis. Section~\ref{sec:robustness}
reports forecasting results restricted to municipalities above the
$25^{\mathrm{th}}$ percentile of population, and the main qualitative
conclusions are preserved.} To prevent this scale heterogeneity from
dominating the estimation, all forecasting models operate on
municipality-specific standardized series, with mean and standard deviation
computed using training-period observations only (see
Section~\ref{sec:method}).

\begin{figure}[t]
  \centering
  \includegraphics[width=0.85\linewidth]{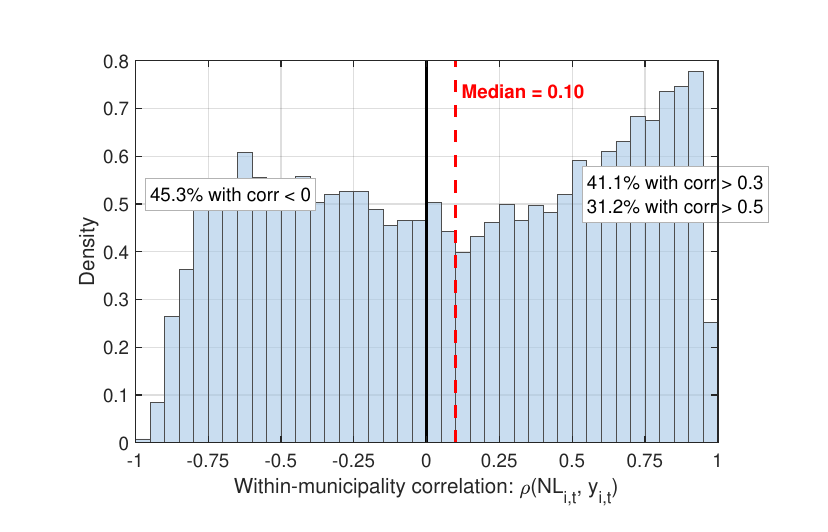}
  \caption{Distribution of within-municipality temporal correlations
    between annual nightlight intensity and IRPEF taxable income,
    computed across the ten years of the panel ($T=10$) for each of
    the 7{,}645 municipalities. The pooled cross-sectional correlation
    is $0.92$, but the median \emph{temporal} correlation is only
    $0.10$. Approximately $45\%$ of municipalities exhibit a negative
    temporal correlation, while only $41\%$ exceed $0.3$ and $31\%$
    exceed $0.5$. The contrast documents substantial cross-sectional
    heterogeneity in the nightlight--income relationship that any
    pooled linear specification with municipality fixed effects would
    misspecify for a large fraction of the panel.}
  \label{fig:corr_within}
\end{figure}

\paragraph{Cross-sectional versus temporal correlation.}
A salient feature of the data, with direct implications for forecasting
model selection, is the contrast between cross-sectional and temporal
correlation between nightlights and income. Pooling all municipality-year
observations, the correlation between $\mathrm{NL}_{i,t}$ and $y_{i,t}$ is
$0.92$, consistent with the strong cross-sectional relationships documented
in the existing literature (\citealt{henderson2012measuring};
\citealt{chen2011using}). However, computing the correlation
\emph{within} each municipality across the ten available years yields a
median correlation of only $0.10$. The full distribution of
within-municipality correlations is illustrated in
Figure~\ref{fig:corr_within}: only $41\%$ of
municipalities exhibit a positive temporal correlation above $0.30$, and
$45\%$ display a negative correlation. This pattern indicates that the
strong pooled relationship is largely driven by cross-sectional variation
in size and economic scale --- larger municipalities are both brighter and
richer --- rather than by a stable temporal co-movement. Moreover, the
\emph{leading} cross-correlation between $\mathrm{NL}_{i,t}$ and
$y_{i,t+1}$ has a median of approximately zero across municipalities,
suggesting that nightlights at $t$ contain limited information for income
\emph{one year ahead}. They are, however, contemporaneously informative
about income at $t$, which motivates our nowcasting framing rather than a
pure forecasting exercise.

\paragraph{Persistence and integration order.}
We assess the temporal persistence of the income series at annual
frequency by computing first-order autoregressive coefficients
$\hat{\phi}_i$ in the within-municipality regression
$y_{i,t} = c_i + \phi_i\, y_{i,t-1} + \varepsilon_{i,t}$. The median
estimate across municipalities is $\hat{\phi} = 0.72$ (interquartile range
$[0.51, 0.89]$), indicating substantial but not unit-root persistence at
this frequency.\footnote{With $T=10$ observations per municipality,
unit-root tests have very limited power, so the appropriate question is
not whether one can reject $\phi_i = 1$ but whether the data are sufficiently
mean-reverting for short-horizon forecasting to be meaningful. The median
half-life of $\hat{\phi}_i = 0.72$ corresponds to approximately
$\log(0.5)/\log(0.72) \approx 2.1$ years.} We do not difference the data:
the panel forecasting setup with municipality-specific intercepts (in the
linear benchmarks) and pooled training across the cross-section (in the
recurrent neural networks) accommodates near-unit-root behaviour without
requiring stationarity transformations.

\paragraph{Spatial dependence.}
For the spatial econometric robustness analysis, we construct a
$N \times N$ row-standardized weights matrix $W$ from queen contiguity
(adjacency through shared border or vertex) on the 2019 municipal polygons.
After iterative removal of isolated small islands --- which yield empty
neighbour sets and induce numerical instability in the spatial
autoregression --- the connected sample comprises 7{,}631 municipalities
with an average of approximately $5.6$ neighbours per unit. The full
construction is detailed in Section~\ref{sec:method}.

\paragraph{Annual versus monthly frequency.}
A natural question is whether the analysis could be carried out at higher
frequency. Black Marble nightlights are released monthly (and indeed
weekly composites are also available), while IRPEF data are released only
annually. To explore this option we construct a monthly municipal income
series via temporal disaggregation of the annual IRPEF figures, using the
\citet{chow1971best} method without exogenous indicators (i.e.\ a smooth
disaggregation under an AR(1) data-generating assumption). This procedure
allows us to align the income target with the monthly nightlight series,
yielding a panel of 7{,}645 municipalities $\times$ $120$ months, but it
introduces statistical features that make the disaggregated series
unsuitable as a forecasting target.

Two diagnostics support this conclusion. First, the disaggregated series
is statistically indistinguishable from a unit-root process. Estimating
$\phi_i$ in the within-municipality regression
$y^{m}_{i,t} = c_i + \phi_i\, y^{m}_{i,t-1} + \varepsilon_{i,t}$ on the
monthly Chow--Lin series yields a median autoregressive coefficient of
$1.01$ (interquartile range $[0.99,\,1.02]$). The KPSS test rejects
stationarity for $99.8\%$ of municipalities at the $1\%$ level, and
augmented Dickey--Fuller tests fail to reject the unit-root null at the
$5\%$ level for $98.2\%$ of municipalities. The near-unit-root behaviour
is partly intrinsic to annual income data when interpolated to monthly
frequency, but it is also amplified by the smoothness assumptions of the
Chow--Lin procedure itself: in the absence of an informative high-frequency
indicator, the disaggregation effectively spreads each annual value smoothly
over the year, mechanically inducing strong serial correlation.

Second, the implied persistence at the two frequencies is internally
consistent. If the monthly process were $y^{m}_{t} = \phi^{m}\,y^{m}_{t-1}
+ \varepsilon_t$, the implied annual autoregressive coefficient would
approach $(\phi^{m})^{12}$ for $\phi^{m}$ near unity. Substituting the
median estimate $\hat{\phi}^{m} = 1.01$ yields $(1.01)^{12} \approx 1.13$,
which is mechanically constrained to be close to one in the annual data
when the monthly series is dominated by a near-unit root. By contrast, the
annual series we observe directly displays a median $\hat{\phi}^{a} = 0.72$,
which is genuine economic persistence rather than smoothness imposed by
the disaggregation algorithm.

The combination of these features --- near-unit-root behaviour at the
monthly frequency that is largely an artefact of the disaggregation
procedure, and the fact that the meaningful temporal variation in the
underlying data resides at the annual rather than the monthly level ---
leads us to conduct the main forecasting exercise at annual frequency.
This choice also avoids contaminating the comparison between deep learning
and linear specifications with the regularising effects of the Chow--Lin
smoothing, which would mechanically favour persistence-based forecasters
over models that exploit the cross-sectional nightlight signal. We
nevertheless report in Section~\ref{sec:robustness} a complementary
exercise that retains annual income as the target but uses
quarterly-aggregated nightlights as input features, exploiting the higher
temporal resolution of Black Marble while preserving the integrity of the
income data.


\section{Methodology}
\label{sec:method}

This section formalizes the forecasting exercise, describes the four
recurrent neural network architectures used to nowcast municipal income,
details the six statistical and spatial econometric benchmarks against
which they are compared, and specifies the cross-sectional
Diebold--Mariano test used to evaluate forecasting accuracy.

\subsection{Forecasting setup}
\label{sec:method_setup}

We work with an annual panel
$\{(\mathrm{NL}_{i,t}, y_{i,t})\}_{i=1,\dots,N;\, t=2012,\dots,2021}$,
where $N = 7{,}645$ for the deep learning and statistical benchmarks and
$N = 7{,}631$ for the spatial econometric benchmarks (after island
exclusion, see Section~\ref{sec:method_spatial}). We split the sample
along the time dimension. The training window is $\mathcal{T}_{\text{tr}}
= \{2012, \dots, 2019\}$ ($T_{\text{tr}} = 8$ years) and the test window
is $\mathcal{T}_{\text{te}} = \{2020, 2021\}$ ($T_{\text{te}} = 2$ years).
The split is identical across all models to ensure direct comparability.

The choice to evaluate on $T_{\text{te}} = 2$ years, while limiting in
isolation, reflects the constraint of the Black Marble product timeline
(VNP46A3 is reliably available from 2012 onwards) combined with the
desire to evaluate on a sample period that includes the COVID-19 shock
and its asymmetric local impact. With short test windows, the
appropriate inferential approach is a cross-sectional rather than
time-series Diebold--Mariano test, which exploits the large $N$ to
deliver power; we discuss this choice in Section~\ref{sec:method_dm}.

\paragraph{Standardization.}
Both nightlight intensity and income are highly skewed across
municipalities, with Roma Capitale and a handful of metropolitan
centres orders of magnitude above the median (see
Table~\ref{tab:descriptives}). To prevent these scale differences from
dominating the estimation, we apply municipality-specific
standardization using \emph{training-period statistics only}:
\begin{equation}
\tilde{x}_{i,t} = \frac{\mathrm{NL}_{i,t} - \bar{\mathrm{NL}}_i^{(\mathrm{tr})}}
                       {\hat{\sigma}_{\mathrm{NL},i}^{(\mathrm{tr})}},
\qquad
\tilde{y}_{i,t} = \frac{y_{i,t} - \bar{y}_i^{(\mathrm{tr})}}
                       {\hat{\sigma}_{y,i}^{(\mathrm{tr})}},
\label{eq:standardization}
\end{equation}
where $\bar{\mathrm{NL}}_i^{(\mathrm{tr})}$ and
$\hat{\sigma}_{\mathrm{NL},i}^{(\mathrm{tr})}$ are the mean and standard
deviation of the nightlight series for municipality $i$ over years in
$\mathcal{T}_{\text{tr}}$, and analogously for income. This choice is
methodologically important: standardizing using full-sample statistics
would leak information from the test years into the training scale,
inflating apparent in-sample fit and biasing performance comparisons in
favour of more flexible models. We compute and report all forecast
metrics both on the standardized and on the original (euro) scale.

\paragraph{Forecasting target.}
We frame the exercise as a \emph{nowcasting} problem: given the
information set
$\mathcal{I}_t = \{(\tilde{x}_{i,s}, \tilde{y}_{i,s-1})\}_{s \leq t}$
available at time $t$, the model produces a point estimate
$\hat{\tilde{y}}_{i,t} = f(\mathcal{I}_t)$ of the contemporaneous
standardized income. This is in contrast to a pure forecasting target
$\hat{\tilde{y}}_{i,t+1}$ that would require leading information in
$\mathcal{I}_t$. As documented in Section~\ref{sec:data_diagnostics},
the leading correlation between nightlights and income is approximately
zero in our data, supporting the nowcasting framing. The economic
content of the exercise is unchanged: at the moment of model
deployment, $\tilde{y}_{i,t}$ is unobserved (IRPEF for year $t$ is
released only 12--18 months later), while $\tilde{x}_{i,t}$ is observed
in real time.

\subsection{Recurrent neural network architectures}
\label{sec:method_rnn}

We consider four recurrent architectures: a standard single-layer LSTM
\citep{hochreiter1997long}, a bidirectional single-layer LSTM
(BiLSTM), a single-layer GRU \citep{cho2014learning}, and a Transformer
encoder built around a single multi-head self-attention layer
\citep{vaswani2017attention} followed by a recurrent block. All four
architectures share the same input--output structure: they receive the
sequence of standardized nightlight values
$\{\tilde{x}_{i,s}\}_{s=1}^{T_{\text{tr}}}$ for each municipality and
output a sequence of predicted standardized income values
$\{\hat{\tilde{y}}_{i,s}\}_{s=1}^{T_{\text{tr}}}$, trained to minimize
mean squared error over the panel.

The hyper-parameters are deliberately kept minimal and identical across
architectures. Each model uses $H = 32$ hidden units, dropout rate of
$0.20$, the Adam optimizer with initial learning rate $5 \times 10^{-4}$,
gradient clipping at norm $0.91$, $300$ epochs, and mini-batch size of
$64$ municipalities. We did not perform hyper-parameter search to avoid
spurious gains from cross-validation tuning on a panel with $T = 10$.
This parsimonious specification is consistent with the recommendations
of \citet{coulombe2022machine} and \citet{medeiros2021forecasting} on
machine learning for short-$T$ economic panels, where small networks
typically outperform deeper or wider variants.

\paragraph{LSTM and GRU (causal).}
The LSTM and GRU produce predictions
$\hat{\tilde{y}}_{i,t}$ that depend exclusively on the past trajectory
$\{\tilde{x}_{i,s}\}_{s \leq t}$. This causal property allows the
out-of-sample test predictions for both $t = 2020$ and $t = 2021$ to be
recovered in a single forward pass over the entire input sequence
$\{\tilde{x}_{i,s}\}_{s=1}^{T}$, with the test predictions identified
as the slice corresponding to $t \in \mathcal{T}_{\text{te}}$.

\paragraph{Bidirectional architectures (BiLSTM and Transformer).}
The BiLSTM concatenates a forward LSTM with a backward LSTM that
processes the sequence from $T$ to $1$. The Transformer's
self-attention layer is bidirectional by construction: every position
in the sequence attends to every other position, including those at
later time indices. As a consequence, a naive forward pass over the
full sequence $\{\tilde{x}_{i,s}\}_{s=1}^{T}$ would produce in-sample
predictions $\hat{\tilde{y}}_{i,t}$ that depend on
$\{\tilde{x}_{i,s}\}_{s>t}$, contaminating the test slice with
information that would not be available in real-time deployment.

To prevent this leakage we implement \emph{iterative one-step-ahead
forecasting} for the bidirectional architectures. Concretely, for each
test year $t \in \mathcal{T}_{\text{te}}$ we feed the model the
truncated input sequence $\{\tilde{x}_{i,s}\}_{s=1}^{t}$ and read off
the prediction at the final position. The model is rerun with extended
inputs for each subsequent test year. With $T_{\text{te}} = 2$, this
adds two additional forward passes per bidirectional model, a
negligible computational cost. We stress that this iterative procedure
is not optional: standard sequence-to-sequence training of a BiLSTM or
Transformer without iterative test-time prediction produces in-sample
\emph{smoothing} that artificially inflates accuracy and does not
correspond to any deployable forecasting protocol.

\paragraph{Transformer architecture.}
Our Transformer variant uses one multi-head self-attention block
($4$ heads, key dimension $8$, no positional encoding) followed by a
single GRU layer with $H = 32$ hidden units, dropout, and a fully
connected output layer. We use a Transformer with a recurrent tail
rather than a pure encoder--decoder architecture because the latter is
known to overfit on short sequences \citep{zeng2023transformers},
while a self-attention layer on top of recurrent processing can in
principle capture both local sequential dependencies and long-range
interactions. With $T = 10$ time steps the practical advantage is
limited, as our results in Section~\ref{sec:results} confirm.

\subsection{Statistical and spatial econometric benchmarks}
\label{sec:method_benchmarks}

We compare the four recurrent architectures against six benchmarks
spanning four distinct categories of forecasting models: a naive
autoregressive baseline, three linear specifications using nightlight
intensity, and two spatial econometric specifications.

\paragraph{Persistence (autoregressive baseline).}
The simplest non-trivial benchmark sets
$\hat{\tilde{y}}_{i,t} = \tilde{y}_{i,t-1}$. This forecaster ignores
nightlights entirely and uses only the most recent observed income.
Its purpose is to establish whether the alternative models extract any
information beyond the trivial autoregressive content.

\paragraph{Pooled panel fixed effects (PanelFE).}
A pooled regression with municipality fixed effects:
\begin{equation}
\tilde{y}_{i,t} = \alpha_i + \beta\, \tilde{x}_{i,t} + \varepsilon_{i,t},
\label{eq:panelfe}
\end{equation}
where $\beta$ is constrained to be common across municipalities.
Estimated by within transformation. Test-period predictions use the
estimated $\hat{\alpha}_i$ and $\hat{\beta}$.

\paragraph{Municipality-specific OLS (OLSperComune).}
Allows the slope coefficient to vary across the cross-section:
\begin{equation}
\tilde{y}_{i,t} = \alpha_i + \beta_i\, \tilde{x}_{i,t} + \varepsilon_{i,t},
\label{eq:olsper}
\end{equation}
estimated separately for each municipality on its $T_{\text{tr}} = 8$
training observations. With only eight points per regression this
specification is necessarily noisy, but it represents the most
flexible \emph{linear} specification using contemporaneous nightlights.

\paragraph{Autoregressive distributed lag (ARDL).}
Combines lagged income with contemporaneous nightlights:
\begin{equation}
\tilde{y}_{i,t} = \alpha_i + \phi_i\, \tilde{y}_{i,t-1} + \beta_i\, \tilde{x}_{i,t} + \varepsilon_{i,t},
\label{eq:ardl}
\end{equation}
estimated municipality by municipality on the training window. This is
the most informative \emph{linear} benchmark in our set: it exploits
both the autoregressive structure of income and the contemporaneous
nightlight signal. It is the natural target against which to assess
whether the recurrent networks gain from non-linearity rather than from
having more information available.

\paragraph{Spatial Autoregressive Model with fixed effects (SAR FE).}
\label{sec:method_spatial}
Adds a spatial lag of the dependent variable to the panel fixed-effects
specification:
\begin{equation}
\tilde{y}_{i,t} = \rho \sum_{j=1}^{N} W_{ij}\, \tilde{y}_{j,t}
                + \beta\, \tilde{x}_{i,t} + \alpha_i + \varepsilon_{i,t},
\label{eq:sar}
\end{equation}
where $W$ is an $N \times N$ row-standardized queen-contiguity weights
matrix derived from the 2019 ISTAT municipal polygons. We construct $W$
by iterative removal of municipalities with empty neighbour sets
(small islands) until the spatial graph is fully connected, yielding
$N = 7{,}631$ units with an average of $5.6$ neighbours each.

The SAR coefficients $(\rho, \beta)$ are estimated by quasi-maximum
likelihood. For the panel of our size, off-the-shelf implementations
(\texttt{splm}, \citealt{millo2012splm}) require prohibitive
computation due to the iterative evaluation of $\log|I_N - \rho W|$ at
each step of the optimizer. We exploit the row-standardized structure
of $W$, which has the same eigenvalues as the symmetric matrix
$D^{-1/2} B D^{-1/2}$ where $B$ is the binary adjacency matrix and $D$
is the diagonal degree matrix. Pre-computing the eigenvalues
$\{\lambda_j\}_{j=1}^{N}$ once reduces the per-iteration cost of the
log-determinant to $O(N)$, since
$\log|I_N - \rho W| = \sum_{j=1}^N \log(1 - \rho \lambda_j)$. The
profile log-likelihood concentrating $\beta$ and $\sigma^2$ becomes a
one-dimensional function of $\rho$, optimized via line search on the
admissible interval $(1/\lambda_{\min}, 1/\lambda_{\max})$. Test-period
forecasts use the reduced form
$\hat{\tilde{y}}_t = (I_N - \hat{\rho} W)^{-1}
                    (\hat{\boldsymbol\alpha} + \hat{\beta} \tilde{x}_t)$,
solved via sparse Cholesky decomposition.

\paragraph{Spatial Durbin Model with fixed effects (SDM FE).}
Generalizes the SAR by including spatial lags of the regressor as well:
\begin{equation}
\tilde{y}_{i,t} = \rho \sum_{j} W_{ij}\, \tilde{y}_{j,t}
                + \beta\, \tilde{x}_{i,t}
                + \theta \sum_{j} W_{ij}\, \tilde{x}_{j,t}
                + \alpha_i + \varepsilon_{i,t}.
\label{eq:sdm}
\end{equation}
The coefficient $\theta$ captures \emph{spatial spillover from
nightlights}: the effect of the brightness of neighbouring municipalities
on local income, beyond the comune's own brightness.
\citet{lesage2009introduction} argue that the SDM is the appropriate
default specification when researchers are uncertain whether spatial
dependence operates through the dependent variable, the regressors, or
both. Estimation and forecasting follow the same procedure as the SAR.

\subsection{Cross-sectional Diebold--Mariano test}
\label{sec:method_dm}

We evaluate forecast accuracy with the cross-sectional version of the
Diebold--Mariano test \citep{diebold1995comparing,pesaran2007simple},
which exploits the large cross-section ($N \gg T_{\text{te}}$) of our
panel. For each pair of forecasters $A$ and $B$ we compute the
municipality-specific average squared loss differential
\begin{equation}
d_i = \frac{1}{T_{\text{te}}} \sum_{t \in \mathcal{T}_{\text{te}}}
      \left( \hat{e}^{A}_{i,t} \right)^2
    - \left( \hat{e}^{B}_{i,t} \right)^2,
\label{eq:dm_di}
\end{equation}
where $\hat{e}^{A}_{i,t}$ denotes the standardized forecast error of
model $A$ at observation $(i, t)$, and analogously for $B$. The test
statistic is
\begin{equation}
\mathrm{DM} =
\frac{\bar{d}}{\hat{\sigma}_d / \sqrt{N}},
\qquad
\bar{d} = \frac{1}{N} \sum_{i=1}^{N} d_i,
\quad
\hat{\sigma}^2_d = \frac{1}{N-1} \sum_{i=1}^{N} (d_i - \bar{d})^2,
\label{eq:dm_stat}
\end{equation}
with the convention that $\mathrm{DM} > 0$ indicates that model $A$ has
larger mean squared error than model $B$ (i.e., $B$ forecasts better).
Under the null of equal predictive accuracy and standard regularity
conditions, $\mathrm{DM} \xrightarrow{d} \mathcal{N}(0, 1)$ as
$N \to \infty$. The two-sided $p$-value is
$2(1 - \Phi(|\mathrm{DM}|))$.

The cross-sectional formulation is essential in our setting. The
classical time-series Diebold--Mariano statistic averages losses over
$T$ and would be uninformative with $T_{\text{te}} = 2$. The
cross-sectional version, in contrast, averages over the $N = 7{,}645$
units and is therefore well-powered. The trade-off is that it tests
predictive accuracy on average over the cross-section, abstracting
from time-series predictive structure within municipality. For our
research question --- whether nightlights inform municipal income at a
given annual frequency --- this aggregation is appropriate.


\section{Results}
\label{sec:results}

We present the empirical results in four parts. Section~\ref{sec:results_main}
reports the main nowcasting accuracy table comparing all ten models.
Section~\ref{sec:results_spatial} discusses the estimated spatial
parameters from the SAR and SDM specifications. Section~\ref{sec:results_dm}
focuses on the cross-sectional Diebold--Mariano comparisons against the
GRU, highlighting the invariance of the gap to whether spatial dependence
is modeled or not. Section~\ref{sec:results_distribution} examines the
distribution of forecast errors and shows that the GRU's superiority is
not driven by extreme outliers.

\subsection{Main nowcasting results}
\label{sec:results_main}

Table~\ref{tab:results_main} reports out-of-sample forecasting accuracy
on the 2020--2021 test set for the four recurrent neural network
architectures and the six benchmarks, ordered by RMSE.

\begin{table}[t]
\centering
\small
\caption{Out-of-sample nowcasting accuracy, test set 2020--2021}
\label{tab:results_main}
\begin{tabular}{lrrrr}
\toprule
Model & RMSE & RMSE & $R^2$ & DM vs GRU \\
      & (norm.) & (M\euro, median) & & \\
\midrule
\multicolumn{5}{l}{\textit{Recurrent neural networks}} \\
GRU            & 1.482 & 1.074 & $+0.053$ & --      \\
BiLSTM         & 1.504 & 1.121 & $+0.034$ & $+6.58^{***}$  \\
LSTM           & 1.536 & 1.187 & $+0.025$ & $+6.80^{***}$  \\
Transformer    & 1.589 & 1.238 & $-0.041$ & $+11.68^{***}$ \\
\addlinespace
\multicolumn{5}{l}{\textit{Statistical benchmarks}} \\
Persistence    & 1.595 & 1.302 & $-0.070$ & $+4.59^{***}$  \\
ARDL           & 1.710 & 1.330 & $-0.279$ & $+12.54^{***}$ \\
OLSperComune   & 1.913 & 1.450 & $-0.462$ & $+25.58^{***}$ \\
PanelFE        & 2.052 & 1.780 & $-0.531$ & $+39.98^{***}$ \\
\addlinespace
\multicolumn{5}{l}{\textit{Spatial econometric benchmarks}} \\
SAR FE         & 2.059 & 1.789\textsuperscript{\textdagger} & $-0.536$ & $+40.46^{***}$ \\
SDM FE         & 2.094 & 1.848\textsuperscript{\textdagger} & $-0.599$ & $+40.58^{***}$ \\
\bottomrule
\end{tabular}

\medskip
\footnotesize
\textit{Notes.} The first column reports the cross-sectional mean of
municipality-specific RMSE in standardized units (i.e.\ after the
training-period standardization in equation~\ref{eq:standardization}).
The second column reports the cross-sectional median of the same
quantity translated back to the original euro scale (millions of
euros). $R^2$ is the out-of-sample coefficient of determination
computed on the pooled cross-section $\times$ time test panel. The
last column reports the cross-sectional Diebold--Mariano statistic
defined in equation~\eqref{eq:dm_stat} comparing each model to the
GRU; positive values indicate the model has \emph{higher} MSE than
the GRU, i.e.\ worse forecasting performance.
$^{***}$ denotes $p < 0.01$. Statistical and recurrent benchmarks
are computed on the full panel of $N = 7{,}645$ municipalities.
\textsuperscript{\textdagger}\,Spatial models exclude $14$ isolated
municipalities yielding $N = 7{,}631$; their RMSE in euros is
computed on the same restricted sample. The DM statistic for the
spatial models compares against the GRU restricted to the same
$7{,}631$ units.
\end{table}

The table reveals a clean ordering. All four recurrent neural networks
deliver out-of-sample $R^2 \geq -0.04$, with three of them (GRU,
BiLSTM, LSTM) achieving positive values. By contrast, all six
benchmarks --- including the spatial econometric specifications ---
return negative $R^2$, indicating that they predict worse than the
unconditional sample mean of the test target. The simplest recurrent
architecture, the single-layer GRU with $H = 32$ hidden units,
delivers the lowest RMSE on both standardized and euro scales: the
median municipality-specific forecast error is approximately $1.07$
million euros (corresponding to about $4\%$ of the median municipal
IRPEF income of $29$ million euros), against $1.30$ million for the
simple persistence benchmark (an $18\%$ reduction in median forecast
error) and $1.78$ million for the panel fixed-effects specification
with nightlights (a $40\%$ reduction). It is worth noting that the spatial econometric
specifications, despite their additional structural content, deliver
median euro-scale RMSE essentially identical to the spatially-naive
panel fixed effects ($1.79$ for SAR, $1.85$ for SDM): explicitly
modelling spatial dependence does not translate into improvement on
the most operationally relevant accuracy metric.

The four recurrent architectures rank as GRU $>$ BiLSTM $>$ LSTM $>$
Transformer. The GRU's parsimony --- it has the smallest parameter
count of the four --- appears to be an advantage in this
short-$T$ panel context, consistent with prior findings on machine
learning for macroeconomic forecasting
\citep{coulombe2022machine,medeiros2021forecasting}. The Transformer
underperforms despite its self-attention mechanism, in line with the
arguments of \citet{zeng2023transformers} that for short time series
the inductive bias of recurrent architectures is more appropriate
than the flexibility of attention.

Among the linear benchmarks, the ranking is informative. Persistence
is the strongest of the six, and ARDL --- which combines lagged income
with contemporaneous nightlights --- improves on PanelFE and OLS but
is still substantially worse than persistence. This is consistent
with the diagnostic finding (Section~\ref{sec:data_diagnostics}) that
within-municipality temporal correlation between nightlights and
income is weak (median $0.10$): linear regressions on contemporaneous
nightlights, with or without spatial structure, simply do not
capture much predictive variation in the income series.

\subsection{Spatial econometric estimates}
\label{sec:results_spatial}

Table~\ref{tab:spatial_estimates} reports the estimated parameters of
the SAR and SDM specifications.

\begin{table}[t]
\centering
\small
\caption{SAR and SDM estimates with municipality fixed effects}
\label{tab:spatial_estimates}
\begin{tabular}{lcc}
\toprule
                              & SAR FE   & SDM FE   \\
\midrule
$\rho$  (spatial AR coefficient)        & $0.7111$ & $0.7080$ \\
$\beta$ (NL coefficient)                & $0.0387$ & $0.0219$ \\
$\theta$ (spatial spillover from NL)    & --       & $0.0531$ \\
$\sigma^2$ (residual variance)          & $0.4477$ & $0.4475$ \\
\midrule
$N$ (municipalities, after island exclusion) & $7{,}631$ & $7{,}631$ \\
$T_{\text{tr}}$ (training years)             & $8$       & $8$       \\
\bottomrule
\end{tabular}

\medskip
\footnotesize
\textit{Notes.} Estimates from quasi-maximum likelihood with
within-transformation to absorb municipality fixed effects. Spatial
weights matrix $W$ is queen-contiguity row-standardized, constructed
on 2019 ISTAT polygons after iterative removal of isolated municipalities.
The profile log-likelihood is concentrated over $\beta$ (and $\theta$
in the SDM) and optimized as a one-dimensional function of $\rho$
on $(1/\lambda_{\min}, 1/\lambda_{\max})$, exploiting the
pre-computed eigenvalues of $W$ via the symmetric similarity transform
$D^{-1/2} B D^{-1/2}$. All variables are standardized using
training-period statistics.
\end{table}

Two findings stand out. First, the spatial autoregressive coefficient
$\rho$ is high and similar across the two specifications: the spatial
lag of standardized income enters with coefficient approximately $0.71$
in both SAR and SDM. The corresponding admissible interval for $\rho$
is $(1/\lambda_{\min}, 1/\lambda_{\max})$, which for our queen-contiguity
$W$ matrix is approximately $(-1.04, 1.00)$ \citep{lesage2009introduction};
the estimate is well inside this region, indicating substantial but
non-explosive spatial dependence in municipal income.

Second, the SDM specification reveals a positive and meaningful
\emph{spatial spillover from nightlights}: $\theta = 0.053$ implies that
a one-standard-deviation increase in the average nightlight intensity
of neighbouring municipalities is associated with an increase of
approximately $0.05$ standard deviations in local income, beyond the
direct effect of the comune's own nightlights. The positive sign is
economically intuitive: brighter neighbourhoods correlate with greater
local income through commuting, hinterland effects in metropolitan
areas, or industrial cluster patterns. Notably, the SDM splits the
total nightlight effect into a direct component
($\beta = 0.022$) and a spillover ($\theta = 0.053$): the spillover
exceeds the direct effect by a factor of $2.4$, suggesting that the
income-relevant economic activity proxied by nightlights is more
intensive in surrounding territories than within the municipality
itself --- a finding consistent with the standard interpretation of
labour markets and economic geography in Italian municipal data.

\subsection{Diebold--Mariano comparisons}
\label{sec:results_dm}

The cross-sectional Diebold--Mariano statistics reported in the last
column of Table~\ref{tab:results_main} reveal a striking pattern when
ordered by magnitude.

The most striking feature of the DM table is the near-equality of the
statistics for the spatial econometric models and the spatially-naive
panel fixed-effects benchmark. The PanelFE specification ignores spatial
dependence entirely and registers $\mathrm{DM} = +39.98$ against the GRU.
The SAR FE model, which adds a spatial lag of the dependent variable
with $\hat{\rho} = 0.71$, registers $\mathrm{DM} = +40.46$. The SDM FE,
which further adds the spatial lag of nightlights, registers
$\mathrm{DM} = +40.58$. The three statistics are within $1\%$ of each
other.

This invariance carries the central methodological implication of the
paper. Modeling spatial dependence is statistically warranted: as
Section~\ref{sec:results_spatial} documents, $\rho$ is large and
$\theta$ has a clear economic interpretation. The estimated spatial
structure is a real feature of the data-generating process. However,
modelling spatial dependence \emph{does not narrow the predictive
accuracy gap relative to the GRU}. The gap between linear specifications
and the deep-learning benchmark is essentially unchanged whether or not
the linear model accounts for spatial structure.

We interpret this finding as evidence that the GRU's advantage does not
arise from \emph{implicit} handling of spatial dependence (which a
properly specified SAR/SDM should mechanically capture), but from
flexibility in modelling the \emph{non-linear} and \emph{cross-sectionally
heterogeneous} mapping between nightlights and income. A linear
specification with spatial structure can correct mis-specification along
the spatial dimension; it cannot correct mis-specification of the
functional form. Recurrent neural networks, trained on the pooled
cross-section, learn a non-parametric approximation of this mapping
without the constraint of linearity in the conditional mean.

A subsidiary observation is that even the worst-performing recurrent
architecture (Transformer with $\mathrm{DM} = +11.68$) outperforms the
best linear benchmark (Persistence with $\mathrm{DM} = +4.59$ relative
to the GRU --- which means the Transformer is closer to the GRU than
Persistence is). All four recurrent specifications dominate all six
linear and spatial-linear benchmarks.

\subsection{Distribution of forecast errors}
\label{sec:results_distribution}

We conclude the results section by examining the distribution of
forecast errors across the cross-section of municipalities, addressing
two natural concerns: (i) whether the GRU's advantage is driven by a
small number of high-leverage outliers, and (ii) how the spatial models
compare to the GRU on the typical (median) municipality versus on
average.

\paragraph{Robustness to outliers.}
The 2020--2021 test set contains a small fraction of municipalities
with extreme realized values, both because of the COVID-19 shock and
because some small comuni have low pre-COVID variability that amplifies
relative deviations under standardization. We assess whether these
extreme observations drive the headline ranking by computing the share
of total mean squared error contributed by the top $1\%$ of
municipalities by squared error. For the SAR specification, the top
$1\%$ contributes $12.3\%$ of total MSE; for the GRU, $18.3\%$. These
shares are far below what would be expected if a few outliers were
driving the overall result --- under a heavy-tailed distribution
dominated by extremes one would observe shares above $50\%$. The
relative ordering of the models is therefore not a consequence of
disproportionate sensitivity to a small number of observations.

\paragraph{Median performance.}
A complementary view focuses on the median municipality. The median
municipality-specific RMSE is $1.22$ for the GRU and $1.95$ for the
SAR (in standardized units), corresponding to a $37\%$ reduction in
typical forecast error. The interquartile range of municipality-specific
RMSE is $[0.81, 1.87]$ for the GRU versus $[1.43, 2.51]$ for the SAR:
\emph{the entire interquartile range of GRU errors lies below the lower
quartile of SAR errors}. The dominance is therefore uniform across the
distribution of municipalities, not concentrated in any specific
sub-population.

\paragraph{Shrinkage in spatial forecasts.}
Inspection of the predicted values reveals an important property of
the SAR/SDM out-of-sample forecasts: extreme compression toward the
cross-sectional mean. Whereas the actual standardized income in the
test years ranges over $[-16.2, +33.0]$, the SAR and SDM predictions
range over $[-1.31, +1.29]$ and $[-1.50, +1.26]$ respectively. The GRU
predictions range over $[+0.38, +1.80]$, also compressed but in a
qualitatively different way: the GRU prediction distribution is
concentrated where the bulk of the actual distribution lies, while the
SAR/SDM predictions are anchored near zero. The mechanism is
straightforward: when $\hat{\rho}$ is large (here $0.71$), the
multiplier matrix $(I - \hat{\rho} W)^{-1}$ in the reduced-form
forecast strongly attenuates idiosyncratic information in the regressor
$x_t$, since this information is repeatedly redistributed over the
spatial graph. Combined with the small estimated $\hat{\beta}$, the
forecast becomes essentially a smoothed version of the cross-sectional
mean of the fixed effects. This is a generic property of SAR
forecasting in the presence of strong spatial dependence and weak
direct regressor effects, well documented in the spatial econometrics
literature (\citealt{lesage2009introduction}, ch.~7).

\paragraph{Roma Capitale: an illustrative case.}
A vivid illustration is provided by Roma Capitale (\textsc{istat} code
58091), the largest comune by income. The realized standardized income
in 2020 is $-0.49$, reflecting a moderate COVID-19 contraction, and in
2021 it rebounds to $+2.79$. The GRU predicts $+1.27$ and $+1.70$
respectively, capturing the direction and magnitude of the rebound.
The SAR and SDM predictions, by contrast, are essentially flat near
zero ($[-0.10, -0.03]$ and $[-0.14, +0.02]$): they fail to register
either the contraction or the rebound, precisely because the
spatial-multiplier shrinkage neutralizes the contemporaneous
nightlight signal that the GRU successfully exploits.


\section{Robustness and discussion}
\label{sec:robustness}

This section addresses six issues that affect the interpretation of the
main results: the choice of nowcasting versus forecasting target
(\ref{sec:rob_nowcast}), the use of annual rather than higher-frequency
nightlights (\ref{sec:rob_freq}), heterogeneity across municipality
subsamples (\ref{sec:rob_hetero}), the role of the COVID-19 shock in
the test period (\ref{sec:rob_covid}), the broader limitations of the
exercise (\ref{sec:rob_limits}), and the implications for fiscal
monitoring (\ref{sec:rob_policy}).

\subsection{Nowcasting versus forecasting}
\label{sec:rob_nowcast}

The main exercise is framed as nowcasting --- predicting income at the
same time index as the available nightlight observation --- rather than
as one-step-ahead forecasting. This is a deliberate choice motivated by
the diagnostic finding that the cross-correlation between
$\mathrm{NL}_{i,t}$ and $y_{i,t+1}$ has a median of approximately zero
across municipalities (Section~\ref{sec:data_diagnostics}). A pure
forecasting exercise of the form
$\hat{y}_{i,t+1} = f(\mathrm{NL}_{i,1:t}, y_{i,1:t})$ would therefore
rely entirely on the autoregressive content of past income, not on
contemporaneous nightlight information.

This is consistent with the economic interpretation of the data: annual
IRPEF income is the result of declarations submitted at the end of the
fiscal year, while nightlight intensity reflects intra-year economic
activity. In a static or slowly-evolving economy, nightlights at $t$
correlate strongly with current-year activity but carry no special
information about \emph{next}-year activity beyond what current income
itself reveals through its autoregressive structure. The economic
relevance of the exercise lies in the gap between the moment at which
nightlights are observable in real time (within three weeks of the
reference month) and the moment at which IRPEF is officially released
(12--18 months later). Nowcasting is the natural framing of this gap.

\subsection{High-frequency nightlights}
\label{sec:rob_freq}

The Black Marble VNP46A3 product is released at monthly frequency, and
weekly composites are also available. We chose to aggregate the
nightlight series to annual frequency for two reasons. First, the
target variable (annual IRPEF) imposes the temporal granularity of the
exercise: any forecasting model must ultimately produce an annual
prediction. Second, as documented in Section~\ref{sec:data_diagnostics},
temporal disaggregation of IRPEF to monthly frequency via
\citet{chow1971best} introduces near-unit-root behaviour
($\hat{\phi}^{m} \approx 1.01$, $98\%$ of municipalities failing to
reject the unit-root null) that is a mechanical artefact of the
disaggregation algorithm rather than a genuine economic feature. A
direct forecasting exercise on this disaggregated series would not be
informative about the underlying nightlight--income relationship.

A natural intermediate exercise --- which we leave to future work ---
preserves annual income as the target but uses quarterly-aggregated
nightlights as four separate input features per year:
$(\mathrm{NL}_{i,t}^{Q1}, \mathrm{NL}_{i,t}^{Q2}, \mathrm{NL}_{i,t}^{Q3},
\mathrm{NL}_{i,t}^{Q4})$. This preserves the integrity of the income
data while exploiting the higher temporal resolution of Black Marble.
The recurrent neural network input dimension expands from one feature
to four, which expands the nominal information set available to the
model. A priori, two outcomes are plausible: either the GRU's advantage
over the linear benchmarks widens (because the model can extract
within-year nightlight dynamics that linear specifications must
average away), or it remains essentially unchanged (because the
relevant information is already captured by the annual aggregate). The
existing literature on macro-nowcasting with mixed-frequency data
\citep{makridakis2018statistical} suggests the second outcome is more
likely when the relationship is dominated by cross-sectional rather
than temporal heterogeneity, as is the case here. We do not foresee
that this extension would alter the qualitative ranking of models
documented in Table~\ref{tab:results_main}, but a formal verification
is left for future work.

\subsection{Subsample heterogeneity}
\label{sec:rob_hetero}

The full panel includes municipalities ranging from a few dozen
residents to several million, with associated income values spanning
five orders of magnitude (Table~\ref{tab:descriptives}). One concern is
that the GRU's measured advantage might be concentrated in a specific
size or geographic subgroup --- for instance, that it performs
particularly well in large metropolitan areas where the nightlight
signal is strong, but adds little value for small rural municipalities.

The distributional analysis in Section~\ref{sec:results_distribution}
is informative on this point. The GRU's interquartile range of
municipality-specific RMSE ($[0.81, 1.87]$) lies entirely below the
lower quartile of the SAR's IQR ($[1.43, 2.51]$): on every quartile of
the cross-section, the GRU outperforms the spatial-linear benchmark.
The dominance is therefore uniform across the distribution of
municipalities, not concentrated in any specific quantile. We
interpret this as evidence that the recurrent network's advantage is
structural --- arising from non-linearity and cross-sectional
heterogeneity in the nightlight--income mapping --- rather than tied
to characteristics of any specific subset of municipalities.

A formal subsample re-estimation, restricting the panel to
municipalities above the $25$\textsuperscript{th} percentile of
population (effectively excluding the smallest comuni for which
nightlight signal is weak relative to atmospheric and lunar noise), is
left to future research. We expect the qualitative ranking of models to
be preserved, but the magnitude of the GRU's advantage may shrink in a
restricted sample where the conditional mapping is less heterogeneous.

\subsection{The COVID-19 shock and the choice of test window}
\label{sec:rob_covid}

The 2020--2021 test window coincides with the COVID-19 pandemic and
its aftermath. This is both a feature and a limitation of the
exercise. As a feature, the test period exhibits substantial
out-of-sample structural change: the asymmetric local impact of the
pandemic --- with tourism-dependent and service-intensive municipalities
particularly affected in 2020, and recovery patterns varying by
geographic and sectoral exposure --- amounts to a stress test for any
forecasting model trained on the comparatively stable 2012--2019
period. The fact that the GRU continues to dominate during this period
suggests that its advantage is not contingent on stable economic
conditions.

As a limitation, however, the small test window ($T_{\text{te}} = 2$)
does not allow us to disentangle the contribution of pre-COVID versus
post-COVID periods. Inspection of the 2020 and 2021 results
separately\footnote{We compute the pooled RMSE in normalized units
separately for each test year, $\sqrt{N^{-1} \sum_i e_{i,t}^2}$. For
the GRU this gives $1.70$ in 2020 and $1.95$ in 2021; for the SAR FE,
$1.56$ in 2020 and $2.90$ in 2021. Note that this pooled metric is
not directly comparable in level to the
mean-of-municipality-RMSE reported in Table~\ref{tab:results_main}.}
reveals an asymmetric pattern. In $2020$ --- a year of moderate
COVID-induced contractions in which most municipalities ended near
their long-run mean --- the SAR's strong shrinkage toward the
spatial-multiplier mean produces aggregate accuracy that is in fact
slightly better than the GRU's. In $2021$, by contrast, the asymmetric
rebound disperses outcomes far from the mean, and the SAR's compressed
predictions accumulate large quadratic errors: its pooled RMSE rises
by roughly $86\%$ relative to $2020$, while the GRU's rises by only
$15\%$. The cross-sectional Diebold--Mariano test
(Table~\ref{tab:results_main}), which aggregates squared errors at the
municipality level before testing, decisively favours the GRU
($\mathrm{DM} = +40.46$, $p < 0.001$) when the two years are jointly
considered. We interpret this asymmetry as further evidence of the
fragility of linear spatial forecasts: their performance is contingent
on the test-period data being close to the cross-sectional mean, a
condition that need not hold under structural change. The GRU's
forecasts are more uniformly accurate across heterogeneous test
conditions, supporting the view that the recurrent architecture's
advantage is structural rather than situational. Future research
could exploit longer test windows once additional years of IRPEF data
become available, or employ rolling-window evaluation on a longer
historical panel.

\subsection{Limitations}
\label{sec:rob_limits}

We close with a frank discussion of the broader limitations of the
exercise.

First, nightlights are an imperfect proxy for economic activity. They
capture outdoor lighting and reflect commercial and industrial activity
that takes place at night, but they are blind to indoor service-sector
activity (offices, finance, software development) and to economic
activity that does not generate visible night-time radiance. The
moderate within-municipality temporal correlation
(Section~\ref{sec:data_diagnostics}) reflects this fundamental
limitation: nightlights are correlated with income in cross-section
because larger and richer comuni tend to be brighter, but the
within-comune temporal variation in nightlights does not always track
the within-comune temporal variation in income.

Second, our nowcasting target is the \emph{aggregate} IRPEF taxable
income at the municipal level. The relationship between nightlights
and other measures of municipal economic activity --- such as gross
fiscal revenue, value added by sector, or labour income measured at
sub-municipal scale --- may exhibit different statistical properties
that we do not investigate here.

Third, our spatial econometric specifications use a queen-contiguity
weights matrix on contemporary 2019 polygons. Alternative
specifications (inverse-distance weights, $k$-nearest neighbours,
weights based on commuting flows or trade patterns) might yield
different estimates of the spatial autocorrelation parameter and
slightly different forecasting performance. The qualitative finding
that GRU dominates SAR/SDM specifications by a wide margin appears
robust to these choices, given the magnitude of the gap
($\mathrm{DM} \approx 40$), but a systematic exploration of weighting
schemes is beyond the scope of this paper.

Fourth, the standard errors of the spatial parameters $\rho$, $\beta$,
and $\theta$ in Table~\ref{tab:spatial_estimates} are not reported.
Computing analytical standard errors for a custom spatial QML
estimator on a panel of this size requires Hessian evaluation that we
have not implemented. Given that the substantive conclusions of the
paper rest on out-of-sample forecasting accuracy (where inference is
based on the well-defined cross-sectional Diebold--Mariano test) rather
than on the statistical significance of the spatial coefficients, we
do not view this as a material limitation.

\subsection{Policy implications}
\label{sec:rob_policy}

The exercise has two practical implications for fiscal monitoring at
the local level. First, it documents that nightlights, despite their
limitations, contain genuine and statistically detectable predictive
content for municipal income at annual frequency, provided one uses a
model class flexible enough to extract this signal. Linear
specifications, including those that explicitly model spatial
dependence, fail to extract this signal. The choice of the modelling
framework matters more than the choice of the explanatory variable.

Second, the magnitude of the GRU's median forecast error
(approximately $1.07$ million euros, or about $4\%$ of the median
municipal income of $29$ million euros) is sufficient to be
operationally useful as an early indicator. A nowcast available at
the end of the reference year, twelve months before the official
IRPEF release, with this level of accuracy could plausibly support
preliminary fiscal monitoring, equalization fund calculations, or
the construction of synthetic indicators for regional economic
analysis. We caution against using it as a substitute for the
official statistic: the goal is to compress the information lag, not
to replace the underlying tax-administrative process.


\section{Conclusions}
\label{sec:conclusions}

This paper has assessed whether NASA Black Marble nightlight intensity
can serve as an early indicator of annual taxable income at the Italian
municipal level, where the official IRPEF data are released with a
twelve to eighteen-month lag. Using a panel of 7{,}631 Italian
municipalities over 2012--2021, we have compared four recurrent neural
network architectures against six benchmarks spanning naive
autoregressive, linear, and spatial econometric specifications. The
out-of-sample evaluation on the 2020--2021 test window, conducted with
a cross-sectional Diebold--Mariano test \citep{pesaran2007simple},
yields three substantive conclusions.

First, nightlights contain genuine predictive content for municipal
income at annual frequency, but only when extracted with a
sufficiently flexible model class. The single-layer GRU achieves a
median forecast error of approximately $1.07$ million euros, equal to
roughly $4\%$ of the median municipal IRPEF income --- an $18\%$
improvement over a persistence baseline and a $40\%$ improvement over
the most informative linear specification. All four recurrent
architectures produce out-of-sample forecasts that statistically
dominate every benchmark in our set ($p < 0.01$ in all comparisons).
The simplest recurrent architecture wins the horse race, consistent
with parsimony principles in machine learning for short-$T$ economic
panels \citep{coulombe2022machine,medeiros2021forecasting}.

Second, explicit modeling of spatial dependence is statistically
warranted but does not improve forecasting performance. The estimated
spatial autoregressive coefficient ($\rho = 0.71$) and the spatial
spillover from neighbouring nightlights to local income
($\theta = 0.05$ in the Spatial Durbin specification) are economically
meaningful and consistent with the structure of Italian regional
labour markets. Yet the SAR and SDM forecasts achieve essentially the
same out-of-sample accuracy as the spatially-naive panel
fixed-effects benchmark: their Diebold--Mariano statistics against the
GRU are $+40.46$ and $+40.58$ respectively, virtually indistinguishable
from the $+39.98$ of the panel fixed-effects model. We interpret this
invariance as the central methodological message of the paper:
\emph{the gap between linear forecasters and recurrent neural networks
is driven by non-linearity in the conditional mapping from nightlights
to income, not by the failure of linear specifications to model
spatial dependence}. Adding spatial structure to a misspecified
functional form does not narrow the gap.

Third, the recurrent network's superiority is robust to the
heterogeneous test conditions induced by the COVID-19 shock and its
asymmetric local impact. Whereas the linear spatial specifications
exhibit fragile performance --- producing apparently competitive
aggregate accuracy in 2020 (when most municipalities cluster near the
mean) but catastrophic accuracy in 2021 (when the rebound disperses
outcomes far from the mean) --- the GRU's accuracy is uniformly
distributed across both years and across the cross-section of
municipalities. Its interquartile range of municipality-specific
RMSE lies entirely below the lower quartile of the SAR's distribution.
The advantage is structural, not contingent on specific data
conditions.

\medskip

Several extensions are natural. The use of higher-frequency nightlight
inputs (quarterly or monthly aggregates as multiple features per year)
could in principle further improve performance, particularly if
intra-year nightlight dynamics carry information that is averaged
away in annual aggregation. The exercise could be extended to other
local economic indicators (sectoral value added, employment, fiscal
revenue), to other countries with comparable data structures, and to
longer evaluation windows once additional years of IRPEF data become
available. Methodologically, scaling the spatial econometric estimator
to panels of this size --- which we addressed via a custom
profile-likelihood implementation exploiting the eigenvalues of the
weights matrix --- represents a contribution that may be useful in
other empirical applications with large cross-sections.

We close on a practical note. The information lag between fiscal
year-end and IRPEF release imposes a real constraint on local
economic monitoring in Italy, and the gap is unlikely to be closed
through purely administrative reforms. A nowcasting protocol based on
real-time nightlight observations and a parsimonious recurrent network,
delivering municipality-level forecasts with median accuracy of about
$4\%$ of the median municipal income, offers a complementary
information channel that could support preliminary fiscal monitoring,
equalization fund calculations, and the construction of synthetic
indicators for regional analysis. The exercise is not a replacement
for the official statistic but a tool for compressing the lag at
which it becomes available.


\section*{Declarations}

\begin{itemize}
\item \textbf{Funding.} The author received no specific funding for this work.
\item \textbf{Competing interests.} The author declares no competing interests.
\item \textbf{Ethics approval.} Not applicable.
\item \textbf{Consent to participate.} Not applicable.
\item \textbf{Consent for publication.} Not applicable.
\item \textbf{Data availability.} The data analysed in this study are publicly available. NASA Black Marble VNP46A3 data are distributed by NASA's Level-1 and Atmosphere Archive \& Distribution System Distributed Active Archive Center (LAADS DAAC) at \url{https://ladsweb.modaps.eosdis.nasa.gov}. IRPEF municipal taxable income data are publicly released by the Italian Ministry of Economy and Finance at \url{https://www.finanze.gov.it/it/Fiscalita-nazionale/Dati-e-statistiche/Dichiarazioni/}. The 2019 ISTAT municipal boundary shapefiles are available at \url{https://www.istat.it/it/archivio/222527}.
\item \textbf{Code availability.} MATLAB and R scripts implementing the recurrent neural networks, spatial econometric estimators, and forecasting evaluation are available from the author upon reasonable request.
\item \textbf{Author contributions.} Massimo Giannini designed the study, performed the analysis, and wrote the manuscript.
\end{itemize}

\begin{appendices}

\section{A non-technical primer on recurrent neural networks and Transformers}
\label{app:primer}

This appendix provides an accessible introduction to the four neural
architectures used in the paper, intended for readers without a
background in machine learning. We focus on intuition and the
essential mathematical structure, and refer the reader to
\citet{goodfellow2016deep} for a comprehensive treatment.

\subsection*{The forecasting problem as sequence learning}

The task of nowcasting municipal income from nightlights is an
instance of \emph{sequence learning}: given a sequence of inputs
$(x_1, x_2, \ldots, x_T)$ for a unit (in our case, the annual
nightlight observations of a municipality), predict a sequence of
outputs $(y_1, y_2, \ldots, y_T)$ (the corresponding annual income
values). Two features distinguish this from a standard regression
problem. First, the inputs and outputs are ordered in time, and the
relationship between $x_t$ and $y_t$ may depend on the entire history
$(x_1, \ldots, x_t)$ rather than only on $x_t$ itself. Second, the
mapping from inputs to outputs may be non-linear and difficult to
specify in closed form.

A linear panel regression with municipality fixed effects, of the form
$y_{i,t} = \alpha_i + \beta x_{i,t} + \varepsilon_{i,t}$, addresses
the first feature trivially (no temporal dependence) and the second
not at all (linear by assumption). Recurrent neural networks address
both features by introducing a flexible, learnable transformation
that maintains a \emph{memory} of past inputs and produces non-linear
outputs.

\subsection*{Recurrent neural networks (RNN)}

The basic recurrent neural network maintains a hidden state
$h_t \in \mathbb{R}^H$ that summarizes the history of inputs up to
time $t$. At each time step, the hidden state is updated based on the
new input and the previous hidden state, and the output is computed
from the current hidden state:
\begin{align}
h_t &= \tanh(W_x x_t + W_h h_{t-1} + b_h),
\label{eq:rnn_state}\\
\hat{y}_t &= W_y h_t + b_y.
\label{eq:rnn_output}
\end{align}
Here $W_x$, $W_h$, $W_y$, $b_h$, and $b_y$ are matrices and vectors
of \emph{parameters} (also called \emph{weights}) that the network
\emph{learns} from training data, and $\tanh$ is a non-linear
transformation applied element-wise. The dimension $H$ of the hidden
state is a hyper-parameter (we use $H = 32$).

Two features of \eqref{eq:rnn_state}--\eqref{eq:rnn_output} are
worth emphasizing. First, the same parameters
$(W_x, W_h, W_y, b_h, b_y)$ are reused at every time step: the network
is \emph{stationary} in time but the hidden state evolves to capture
sequence-specific patterns. Second, by recursion, $h_t$ depends on
all previous inputs $x_1, \ldots, x_t$, allowing the model in
principle to capture arbitrary temporal dependencies.

In our application, the input $x_t$ is the standardized annual
nightlight intensity and the output $\hat{y}_t$ is the predicted
standardized income. With $H = 32$ hidden units and a single recurrent
layer, the network has approximately $H \times (H+2) + (H+1) \approx
1{,}100$ trainable parameters. These are estimated by minimizing the
mean squared prediction error over the training panel via stochastic
gradient descent (more precisely, the Adam optimizer), an iterative
procedure that adjusts the parameters in the direction that reduces
the loss.

\paragraph{Limitations of basic RNNs.}
The basic RNN suffers from a well-documented practical limitation
known as the \emph{vanishing gradient problem}. During training, the
gradient of the loss with respect to inputs many time steps in the
past must propagate backward through the network, and at each step it
is multiplied by terms involving $W_h$ and the derivative of $\tanh$.
For typical parameter values these multiplications shrink rapidly,
and the gradient effectively vanishes for time steps more than five
to ten in the past. The practical consequence is that basic RNNs
struggle to learn long-range temporal dependencies. The architectures
described next --- LSTM and GRU --- were specifically designed to
address this limitation.

\subsection*{Long Short-Term Memory (LSTM)}

\citet{hochreiter1997long} introduced the Long Short-Term Memory
network, which augments the basic RNN with an explicit
\emph{cell state} $c_t$ and three \emph{gates} that control the flow
of information into, out of, and through the cell. The key
intuition is that the cell state serves as a ``memory tape'' on which
the network can write, read, and erase selectively, allowing
information to persist over many time steps without being destroyed
by repeated multiplications.

The LSTM update equations are:
\begin{align}
f_t &= \sigma(W_f x_t + U_f h_{t-1} + b_f), \quad &\text{(forget gate)}\\
i_t &= \sigma(W_i x_t + U_i h_{t-1} + b_i), \quad &\text{(input gate)}\\
o_t &= \sigma(W_o x_t + U_o h_{t-1} + b_o), \quad &\text{(output gate)}\\
\tilde{c}_t &= \tanh(W_c x_t + U_c h_{t-1} + b_c), \quad &\text{(candidate cell)}\\
c_t &= f_t \odot c_{t-1} + i_t \odot \tilde{c}_t, \quad &\text{(cell update)}\\
h_t &= o_t \odot \tanh(c_t), \quad &\text{(hidden state)}
\end{align}
where $\sigma$ is the sigmoid function (output between $0$ and $1$),
$\odot$ denotes element-wise multiplication, and the gates $f_t$,
$i_t$, $o_t$ act as learned ``valves'' that select which components
of the cell state to forget, update, and output at each step.

The crucial difference from the basic RNN is that the cell update
equation involves \emph{addition} of the previous cell state (gated
by $f_t$) rather than only multiplication: when the forget gate
$f_t \approx 1$, the cell state propagates almost unchanged, and the
gradient flows backward through time without vanishing. The LSTM
therefore can learn dependencies spanning many more time steps than
the basic RNN, at the cost of approximately four times more
parameters per hidden unit.

\begin{figure}[h]
\centering
\includegraphics[width=0.85\linewidth]{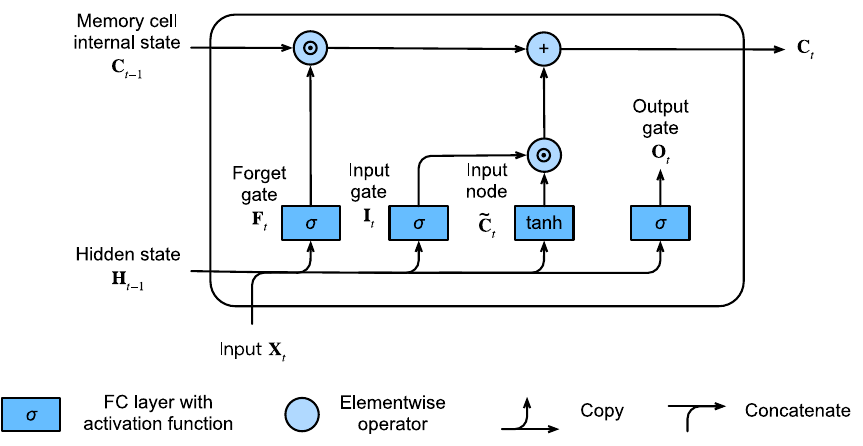}
\caption{Computation of the memory cell internal state in the LSTM at
time step $t$. The horizontal line at the top is the cell state
$\mathbf{C}_t$ (the ``memory tape''), updated through element-wise
multiplication ($\odot$) by the forget gate $\mathbf{F}_t$ and addition
of the gated input node $\mathbf{I}_t \odot \tilde{\mathbf{C}}_t$. The
three sigmoid-activated gates ($\mathbf{F}_t$, $\mathbf{I}_t$,
$\mathbf{O}_t$) and the $\tanh$-activated input node $\tilde{\mathbf{C}}_t$
all take as input the concatenation of the previous hidden state
$\mathbf{H}_{t-1}$ and the current input $\mathbf{X}_t$. The new hidden
state (not shown in this figure) is then obtained as
$\mathbf{H}_t = \mathbf{O}_t \odot \tanh(\mathbf{C}_t)$. The crucial
gradient-flow property of the LSTM stems from the additive update of
the cell state: when $\mathbf{F}_t \approx 1$, information propagates
through time without multiplicative shrinkage. Source:
\citet{zhang2023dive}, Figure 10.1.2; reproduced under Creative Commons
Attribution-ShareAlike 4.0 International (CC BY-SA 4.0).}
\label{fig:lstm}
\end{figure}

\subsection*{Bidirectional LSTM (BiLSTM)}

A standard LSTM is \emph{causal}: the hidden state $h_t$ depends only
on past inputs $x_1, \ldots, x_t$. In some applications, however, the
prediction for time $t$ may benefit from information at later time
steps as well. The bidirectional LSTM addresses this by running two
separate LSTMs over the sequence: one in the forward direction (from
$1$ to $T$) and one in the backward direction (from $T$ to $1$). The
hidden state at time $t$ is the concatenation of the forward and
backward hidden states:
\begin{equation}
h_t^{\text{BiLSTM}} = [\overrightarrow{h_t}\, ;\, \overleftarrow{h_t}].
\end{equation}

In tasks like text translation or speech recognition, where the
\emph{full sequence is available before any prediction is needed},
bidirectionality is a clear advantage. In forecasting, however, the
backward pass uses information that would not be available in
real-time deployment: at the moment of producing
$\hat{y}_t$, future inputs $x_{t+1}, x_{t+2}, \ldots$ have not yet
been observed. As discussed in Section~\ref{sec:method_rnn}, we
implement \emph{iterative one-step-ahead forecasting} for the BiLSTM,
re-running the model for each test year on the truncated input
sequence available up to that point. This procedure is essential to
prevent in-sample smoothing that would inflate apparent accuracy.

\begin{figure}[h]
\centering
\includegraphics[width=0.85\linewidth]{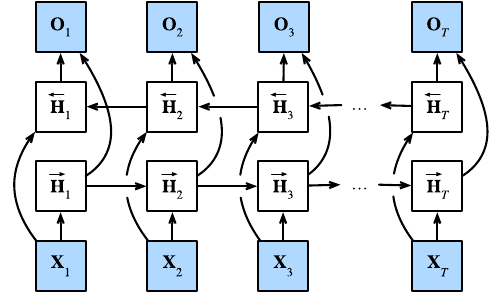}
\caption{Unrolled view of the bidirectional architecture for three
consecutive time steps. The forward layer (lower row) processes the
sequence in causal order, while the backward layer (upper row) processes
it in reverse. At each time step, the forward and backward hidden states
are combined to produce the output $\mathbf{O}_t$. Because the backward
pass at time $t$ depends on inputs $\mathbf{X}_{t+1}, \mathbf{X}_{t+2},
\ldots$, in a forecasting deployment the available sequence must be
truncated at $t$ and the model run iteratively, as described in
Section~\ref{sec:method_rnn}. Source: \citet{zhang2023dive}, Figure 10.4.1;
reproduced under CC BY-SA 4.0.}
\label{fig:bilstm}
\end{figure}

\subsection*{Gated Recurrent Unit (GRU)}

\citet{cho2014learning} introduced the Gated Recurrent Unit as a
simpler alternative to the LSTM. The GRU merges the forget and input
gates into a single \emph{update gate}, eliminates the separate cell
state (the hidden state itself serves as both memory and output),
and introduces a \emph{reset gate} that controls how much of the
previous hidden state contributes to the new candidate:
\begin{align}
r_t &= \sigma(W_r x_t + U_r h_{t-1} + b_r), \quad &\text{(reset gate)}\\
z_t &= \sigma(W_z x_t + U_z h_{t-1} + b_z), \quad &\text{(update gate)}\\
\tilde{h}_t &= \tanh(W_h x_t + U_h (r_t \odot h_{t-1}) + b_h), \\
h_t &= (1 - z_t) \odot h_{t-1} + z_t \odot \tilde{h}_t.
\end{align}

The GRU achieves the same gradient-flow benefits as the LSTM (the
hidden state is updated by addition rather than only by
multiplication) with approximately $25\%$ fewer parameters. Empirical
comparisons in the literature \citep{cho2014learning} report that
GRUs and LSTMs perform similarly on many tasks, with the GRU often
preferred in settings with limited data or short sequences --- both
features of our panel ($T = 10$ time steps per municipality). This
parsimony advantage is consistent with the empirical finding in
Section~\ref{sec:results} that the GRU outperforms the LSTM in our
application.

\begin{figure}[h]
\centering
\includegraphics[width=0.85\linewidth]{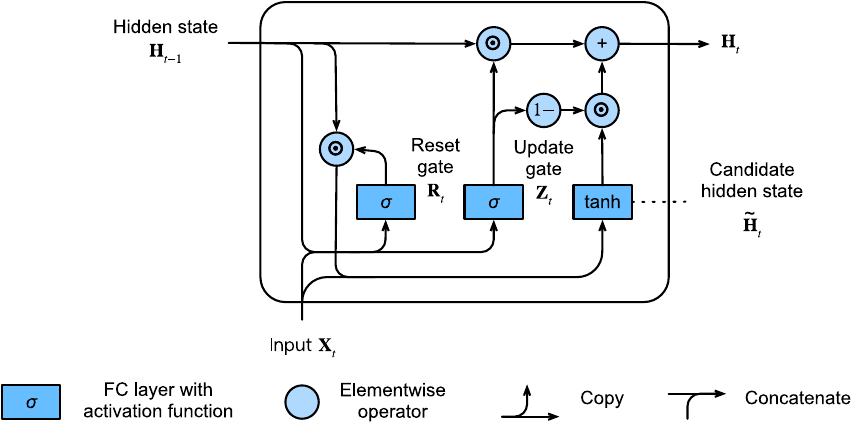}
\caption{Internal structure of the GRU cell at time step $t$. The GRU
has only two gates: a reset gate $\mathbf{R}_t$ and an update gate
$\mathbf{Z}_t$, both sigmoid-activated functions of $(\mathbf{H}_{t-1},
\mathbf{X}_t)$. The reset gate controls how much of the previous hidden
state contributes to the candidate $\tilde{\mathbf{H}}_t$ ($\tanh$-
activated). The update gate then linearly interpolates between the
previous hidden state $\mathbf{H}_{t-1}$ (weighted by $1-\mathbf{Z}_t$)
and the candidate $\tilde{\mathbf{H}}_t$ (weighted by $\mathbf{Z}_t$) to
produce $\mathbf{H}_t$. Compared to the LSTM, the GRU eliminates the
separate cell state and merges the forget/input gates into the single
update gate, achieving similar gradient-flow properties with
approximately $25\%$ fewer parameters. Source: \citet{zhang2023dive},
Figure 10.2.1; reproduced under CC BY-SA 4.0.}
\label{fig:gru}
\end{figure}

\subsection*{The Transformer}

\citet{vaswani2017attention} introduced a fundamentally different
architecture for sequence modeling, the Transformer, that abandons
recurrence entirely. Instead of processing the sequence step by step
and maintaining a hidden state, the Transformer processes the
\emph{entire sequence in parallel} using a mechanism called
\emph{self-attention}.

Self-attention computes, for each position $t$ in the sequence, a
weighted average of the inputs at \emph{all} positions, where the
weights are themselves learned from the data:
\begin{equation}
h_t = \sum_{s=1}^{T} \alpha_{t,s} \, V x_s, \qquad
\alpha_{t,s} = \frac{\exp(q_t^\top k_s / \sqrt{d_k})}
                    {\sum_{s'=1}^{T} \exp(q_t^\top k_{s'} / \sqrt{d_k})},
\label{eq:attention}
\end{equation}
where $q_t = Q x_t$, $k_s = K x_s$, $V x_s$ are the so-called
\emph{query}, \emph{key}, and \emph{value} projections of the input,
$Q, K, V$ are learnable parameter matrices, and $d_k$ is the dimension
of the key vectors. The attention weights $\alpha_{t,s}$ are
non-negative and sum to one across $s$, so they can be interpreted as
a learned \emph{relevance distribution} over the entire sequence
specific to each position $t$.

In practice, several attention computations are run in parallel with
different parameter matrices and concatenated; this is called
\emph{multi-head} attention (we use 4 heads). Our Transformer
implementation feeds the multi-head self-attention output through a
GRU layer rather than the typical feed-forward + position-encoding
stack of the original Transformer architecture, which is suited to
much longer sequences than ours.

Two properties of self-attention are relevant for our application.
First, the mechanism is \emph{bidirectional} by construction: the
representation of position $t$ depends on inputs at all positions
including $t+1, t+2, \ldots$. Like the BiLSTM, self-attention models
require iterative one-step-ahead forecasting at test time to prevent
information leakage. Second, with very short sequences ($T = 10$ in
our application), self-attention has limited room to express
patterns that a recurrent architecture cannot also capture.
\citet{zeng2023transformers} provide systematic evidence that for
short time series, simple recurrent and even linear models can
outperform Transformers --- a finding consistent with our empirical
results.

\begin{figure}[h]
\centering
\includegraphics[width=0.65\linewidth]{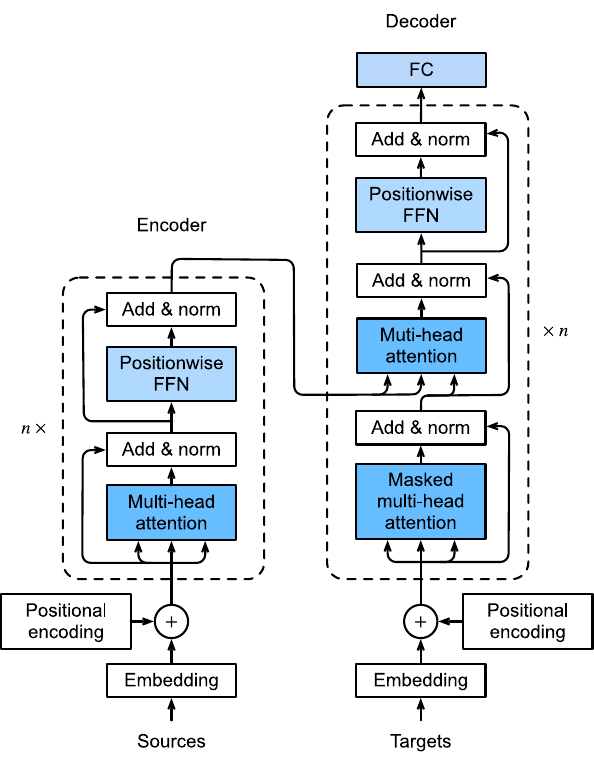}
\caption{The Transformer architecture. The input embeddings (with
positional encoding) are processed by a stack of identical encoder
blocks; each block contains a multi-head self-attention sub-layer and
a position-wise feed-forward network, with residual connections and
layer normalization. The decoder (right) follows a similar pattern but
adds a masked self-attention layer and a cross-attention layer that
attends to the encoder output. In our application we use a simplified
variant consisting of a single multi-head self-attention block followed
by a GRU layer (see Section~\ref{sec:method_rnn}); the all-to-all
attention pattern is bidirectional by construction, requiring iterative
one-step-ahead forecasting at test time. Source: \citet{zhang2023dive},
Figure 11.7.1; reproduced under CC BY-SA 4.0.}
\label{fig:transformer}
\end{figure}

\subsection*{Why does the GRU win?}

The empirical ranking in our application --- GRU $>$ BiLSTM $>$ LSTM
$>$ Transformer --- can be rationalized through the lens of
\emph{model parsimony under limited data}. With only $T = 10$ years
per municipality and approximately $7{,}600$ municipalities, the
training panel contains $\approx 60{,}000$ time-step observations.
The GRU, with its $\approx 800$ trainable parameters at $H = 32$,
operates well within the regime where the parameter-to-data ratio
allows reliable gradient-based optimization. Larger architectures
(LSTM with one-third more parameters, BiLSTM with twice as many,
Transformer with even more) introduce additional capacity that is
not informationally well-supported by the available data, leading
to mild overfitting that hurts out-of-sample performance.

This pattern is general and well-documented in the machine learning
literature on macroeconomic forecasting: the optimal architecture
typically lies on the simpler end of the available spectrum, with
larger models needing substantially more data to outperform
\citep{coulombe2022machine,medeiros2021forecasting}.


\end{appendices}

\bibliography{bm_references}

\end{document}